\documentclass[aps,onecolumn,groupedaddress]{revtex4}

\usepackage{latexsym}
\usepackage{times}
\usepackage{graphicx}
\usepackage{hyperref}
\usepackage{epsfig}

\begin{document}
\thispagestyle{empty}
\pagestyle{empty}
\bibliographystyle{apsrev}

\title{Dispersive Cylindrical Cloaks under Non-Monochromatic Illumination}

\author{Christos Argyropoulos}
\email[]{christos.a@elec.qmul.ac.uk}
\affiliation{Department of Electronic Engineering, Queen Mary, University of London, Mile End Road, London, E1 4NS, United Kingdom}
\author{Efthymios Kallos}
\affiliation{Department of Electronic Engineering, Queen Mary, University of London, Mile End Road, London, E1 4NS, United Kingdom}
\author{Yang Hao}
\email[]{yang.hao@elec.qmul.ac.uk}
\affiliation{Department of Electronic Engineering, Queen Mary, University of London, Mile End Road, London, E1 4NS, United Kingdom}

\date{\today}

\begin{abstract}
Transformation-based cylindrical cloaks and concentrators are illuminated with non-monochromatic waves and unusual effects are observed with interesting potential applications. The transient responses of the devices are studied numerically with the Finite-Difference Time-Domain method and the results are verified with analytical formulas. We compute the effective bandwidth of several cloaking schemes as well as the effect of losses on the performance of the structures. We also find that narrowband behavior, frequency shift effects, time delays and spatial disturbances of the incoming waves are dominant due to the inherently dispersive nature of the devices. These effects are important and should be taken into account when designing metamaterial-based devices.
\end{abstract}


\maketitle

\section{Introduction}
\label{Introduction}

Transformation electromagnetics provides the unprecedented opportunity to manipulate the propagation of electromagnetic waves in an artificial way \cite{Pendry,Leonhardt}. Form-invariant coordinate transformations cause the electromagnetic waves to perceive space compressed or dilated in different coordinate directions, which often requires materials with anisotropic and spatially-varying values of permittivity and permeability tensors. Many interesting applications have been proposed, derived from different coordinate transformations. One of the most widely studied application is the cloak of invisibility \cite{Pendry}, an approximate design of which was constructed for microwaves using arrays of split-ring resonators \cite{Schurig}. Other proposed devices include carpet cloaking structures \cite{JensenLi,LiuR}, ``cloaking'' absorbers \cite{argyropoulos2008mle}, concave mirrors for all angles \cite{yang2008ses}, field concentrators \cite{Rahm}, spherical \cite{Psaltis} and cylindrical \cite{yan2008csc} superlenses, flat near-field and far-field focusing lenses \cite{R.Smith}, reflectionless beam shifters and beam splitters \cite{M.Rahm}, field rotators \cite{H.Chen}, adaptive beam benders and expanders \cite{rahm2008tod} and novel antenna designs \cite{kong91kjh,jiang2008lhg}.

In order to achieve the required material values, the majority of the aforementioned devices are constructed with metamaterials consisting of resonating structures \cite{Robbins,Vier}, whose properties are usually dominated by high loss and frequency dispersion \cite{Wiltshire,podolskiy2005nss}. Any material that exhibits permittivity and permeability values smaller than unity must be dispersive due to causality constraints \cite{tretyakov2007vmp}, and hence, the corresponding devices operate properly only over very narrow bandwidths. One common example of such dispersive materials found in nature is the response of plasma electron gas to external radiation \cite{chen1984ipp}, which exhibits an effective refractive index smaller than one. Consequently, unless such a metamaterial-based device is illuminated by a unrealistic perfect monochromatic wave, a complete description of the response of such devices requires consideration of the full dispersive effects occurring over a range of frequencies. The goal of this paper is to investigate the physics of the broadband response of such dispersive devices.

So far, two-dimensional (2D) cylindrical electromagnetic cloaks have been studied only under monochromatic plane wave illumination \cite{Pendry,Ruan,Cummer,Zhang,Yan}. However, all electromagnetic waves extend over a finite bandwidth. Some dispersive effects have been revealed analytically for three-dimensional (3D) spherical cloaks \cite{zhang2008rab,chen2008tda,zhangoptexpr2009}. For example, blueshift effects have been theoretically predicted \cite{zhang2008rab}, where waves with different frequencies penetrate differently inside the cloaking shell region. Furthermore, peculiar energy transport velocity distributions inside the same device have been predicted \cite{chen2008tda}, using a Hamiltonian optics approach. Recently, the spatial energy distribution of a Gaussian light pulse after passing through such a spherical cloak has been shown to be distorted in a theoretical full-wave analysis \cite{zhangoptexpr2009}. However, these interesting effects have not been verified, until now, with a fully explicit numerical technique and especially for 2D cylindrical cloaks. In addition, the investigation of the cloaking bandwidth, which is constrained by the resonant nature of the metamaterial elements, has been very limited in the literature to mostly analytical treatments \cite{zhang2008rab,Ivsic}.

In this paper, we utilize numerical simulations to examine the physics behind the dispersive and transient nature of 2D lossless cylindrical cloaks. The ideal \cite{Cummer} and matched reduced \cite{Qiu} cylindrical cloaks are illuminated with non-monochromatic Gaussian electromagnetic pulses, and their transient response is analyzed with a radially-dependent dispersive Finite-Difference Time-Domain (FDTD) method \cite{Yan,Argyropoulos}. The time-domain numerical technique presented here is advantageous compared to the Finite Element Method (FEM) used in previous works \cite{Cummer}, since the transient response and the operational bandwidth of a device can be easily computed. First, we take advantage of the time-domain properties of FDTD to compute the bandwidth of some popular cloaking schemes. We show that the cloaking performance is very limited when reasonable losses are included in the model. We also find that, under such non-monochromatic illumination, even the ideal electromagnetic cloak can be detected due to significant frequency shifts that occur because of the dispersion in the cloaking shell. This can be explained by the fact that the behavior of the waves at frequencies below the operating frequency of the cloak, where a negative-index shell may be formed, is fundamentally different than the behavior of waves above that frequency, where the cloak is found to act more like a dielectric scatterer. The Drude dispersion model is used to explain the origin of these results. In addition, variations in the group velocity and energy distribution across dispersive materials are found to cause severe distortion of the recomposed wavefronts after the cylindrical cloak and the electromagnetic concentrator devices. As a result, the incident pulses are distorted both temporally as well as spatially. These findings are not limited to the transformation-based cloaks examined in this work, but are expected to arise in a similar fashion in any dispersive and anisotropic device based on transformation electromagnetics.

The rest of the paper is organized as follows. In Sec. \ref{Radially-Dependent Dispersive FDTD} we present the details of the dispersive FDTD code utilized in the numerical simulations. In Sec. \ref{Bandwidth and Loss Limitation of cylindrical cloaks} the bandwidth and loss limitations of different cylindrical cloaks are explored. In Sec. \ref{Spectral Response of the Ideal Cylindrical Cloak} the dispersive effects in the ideal cylindrical cloak are analyzed by launching non-monochromatic pulses towards the device, where frequency shift effects are examined. In Sec. \ref{Spectral Response of the Reduced Cylindrical Cloak} the corresponding dispersive effects in the reduced cylindrical cloak are investigated. Finally, in Sec. \ref{Temporal and Spatial Responses of the Ideal Cylindrical Cloak and Concentrator} the transient responses of the ideal cylindrical cloak and the ideal concentrator are investigated, which give rise to time-delay effects and non-uniform transmitted spatial energy distributions.

\section{Radially-Dependent Dispersive FDTD}
\label{Radially-Dependent Dispersive FDTD}

\begin{figure}[t]
\centering
\includegraphics[width=8.0cm]{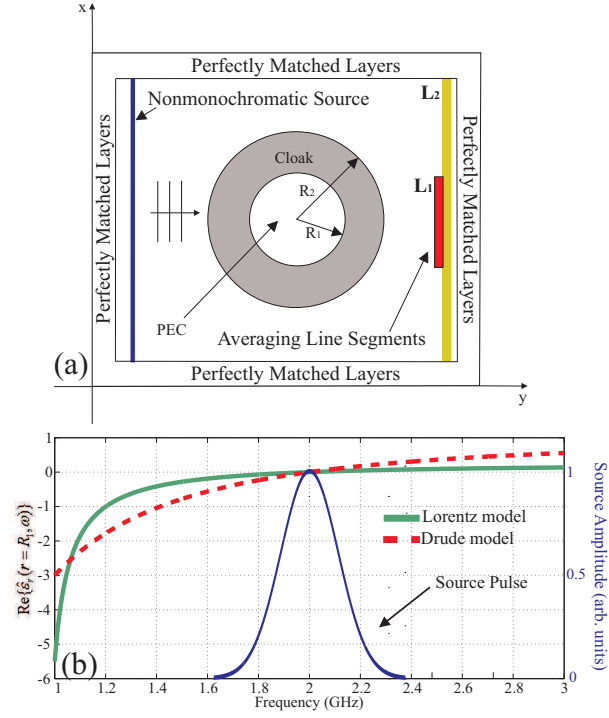}
\caption{(a) FDTD computational domain of the cylindrical cloaking structure for the case of non-monochromatic plane wave illumination. The fields are computed using the Total-Field Scattered-Field (TF-SF) technique \cite{Taflove05}. (b) Comparison between Drude and Lorentz dispersion models, when used to map the $\hat{\varepsilon}_{r}$ parameter at the point $r=R_{1}$. The bandwidth of a typical incident pulse is also plotted in the same graph to illustrate the dispersion along its spectrum.}
\label{domainandblueshift}
\end{figure}

In this section details of the FDTD simulations relevant to the modeling of dispersive devices are presented. For the 2D FDTD simulations, transverse electric (TE) polarized wave incidence is assumed, without loss of generality, where only three field components are non-zero: $E_{x},E_{y}$ and $H_{z}$. The FDTD cell size, throughout the modeling, is chosen $\Delta x=\Delta y=\lambda/150$, where $\lambda$ is the wavelength of the excitation signal in free space. The domain size is $850\times850$ cells, or approximately $5.66\lambda\times5.66\lambda$. The temporal discretization is chosen according to the Courant stability condition \cite{Taflove05} and the time step is given by $\Delta t=\Delta x/\sqrt{2}c$, where $c$ is the speed of light in free space. Throughout the paper the devices are designed to have an operating frequency of $f_{0}=2.0$ GHz, where the free space wavelength is $\lambda\simeq15$ cm. The cloaked object is chosen to be a perfect electric conductor (PEC) material. Unless otherwise noted, the dimensions of the cloaking structure are $R_{1}=\frac{2\lambda}{3}$ and $R_{2}=\frac{4\lambda}{3}$ in terms of the free space wavelength. Here $R_{1}$ is the inner radius (radius of PEC cylinder) and $R_{2}$ the outer radius of the cloaking shell. The FDTD computational domain used throughout this paper is shown in Fig. \ref{domainandblueshift}(a).

Plane waves centered at different frequencies with variable temporal narrowband and broadband Gaussian envelopes are used to illuminate the structures. The $H_{z}$ field amplitude values after the cloak are averaged along two overlapping line segments with different lengths, $L_{1}$ and $L_{2}$, close to the right side of the domain, as can be seen in Fig. \ref{domainandblueshift}(a). These segments are located at the same point along the y-axis, despite being depicted at slightly different locations for clarity purposes. The frequency spectra of the transmitted pulses are then retrieved after recording the time history of the evolution of the field amplitude along these line segments.

In this paper, the two main examples of interest are the ideal cloak and the matched reduced cloak. The parameters of the former device are given in cylindrical coordinates as \cite{Cummer}:
\begin{eqnarray}
\varepsilon_{r}(r)&=&\frac{r-R_{1}}{r}\nonumber\\
\varepsilon_{\phi}(r)&=&\frac{r}{r-R_{1}}\label{parametersofcloaking}\\
\mu_{z}(r)&=&\left(\frac{R_{2}}{R_{2}-R_{1}}\right)^{2}\frac{r-R_{1}}{r}\nonumber
\end{eqnarray}
The parameters of the matched reduced cloak are instead \cite{Qiu}:
\begin{eqnarray}
\varepsilon_{r}(r)&=&\frac{R_{2}}{R_{2}-R_{1}}\left(\frac{r-R_{1}}{r}\right)^{2}\nonumber\\
\varepsilon_{\phi}(r)&=&\frac{R_{2}}{R_{2}-R_{1}}\label{parametersofreducedcloaking}\\
\mu_{z}(r)&=&\frac{R_{2}}{R_{2}-R_{1}}\nonumber
\end{eqnarray}
Here $R_{1}<r<R_{2}$ is an arbitrary radius inside the cloaking shell.

The material parameters of the devices discussed here can have values less than one, as extracted from the coordinate transformations, indicating that the required materials are frequency dispersive. The dispersive region of the parameters is mapped with the Drude model \cite{Stewart} in the FDTD technique. For the results shown in this paper, it should be noted that physically similar results are obtained, when the simulations are repeated using the Lorentz dispersion model \cite{Robbins}. For the case of the radial components of the frequency-dependent relative permittivity $\hat{\varepsilon}_{r}$ and permeability $\hat{\mu}_{r}$ in the FDTD code, they are given by:
\begin{eqnarray}
\hat{\varepsilon}_{r}(r,\omega)=1-\frac{\omega_{p\varepsilon}^2(r)}{\omega^2-\jmath\omega\gamma}
\label{Drudemodel}
\end{eqnarray}
\begin{eqnarray}
\hat{\mu}_{r}(r,\omega)=1-\frac{\omega_{p\mu}^2(r)}{\omega^2-\jmath\omega\gamma}
\label{Drudemodel2}
\end{eqnarray}
where $\omega_{p\varepsilon}$ and $\omega_{p\mu}$ are the plasma frequencies and $\gamma$ is the collision frequency, which is set to zero for the current lossless cloaks. The plasma frequencies are equal to:
\begin{eqnarray}
\omega_{p\varepsilon}(r)=\omega_{0}\sqrt{1-\varepsilon_{r}(r)}
\label{plasmafrequencyanalytical_e}
\end{eqnarray}
\begin{eqnarray}
\omega_{p\mu}(r)=\omega_{0}\sqrt{1-\mu_{r}(r)}
\label{plasmafrequencyanalytical_m}
\end{eqnarray}

Here $\omega_{0}$ is the device's operating frequency, while $\varepsilon_{r}(r)$ and $\mu_{r}(r)$ are the desired frequency-independent material parameters at the design frequency $\omega_{0}$ of the device. Equations (\ref{plasmafrequencyanalytical_e}) and (\ref{plasmafrequencyanalytical_m}) are substituted in Eq. (\ref{Drudemodel}) and the resulted material parameters of the cloak are explicitly obtained:
\begin{eqnarray}
\hat{\varepsilon}_{r}(r,\omega)=1-\frac{\omega_{0}^2}{\omega^2}[1-\varepsilon_{r}(r)]
\label{radandfreqdependparam_e}
\end{eqnarray}
\begin{eqnarray}
\hat{\mu}_{r}(r,\omega)=1-\frac{\omega_{0}^2}{\omega^2}[1-\mu_{r}(r)]
\label{radandfreqdependparam_m}
\end{eqnarray}
Note that the model is build such that the device's design parameters are retrieved exactly when $\omega=\omega_{0}$. In that case $\hat{\varepsilon}_{r}(r,\omega_{0})=\varepsilon_{r}(r)$ and $\hat{\mu}_{r}(r,\omega_{0})=\mu_{r}(r)$.

Figure \ref{domainandblueshift}(b) shows the value of the real part of $\hat{\varepsilon}_{r}$ as a function of frequency at the inner surface of an ideal cylindrical cloak, when either the Drude or Lorentz models are used. As a reference, the spectrum of an incident Gaussian pulse centered at $2.0$ GHz with a bandwidth of $200$ MHz  (Full Width at Half Maximum - FWHM) is also plotted in the same graph, in order to illustrate the variation due to frequency dispersion of the material parameters across a typical pulse. The graph shows that, independently of which dispersive model is used, waves at some frequencies perceive the inner core of the cloak as a negative-permittivity material, while waves at other frequency bands perceive it as a positive-permittivity material. As it shall be shown, this asymmetry in frequency, which is inherent to all dispersive materials, is the fundamental reason that gives rise to the effects that are described in the sections \ref{Spectral Response of the Ideal Cylindrical Cloak} and \ref{Spectral Response of the Reduced Cylindrical Cloak}.

The devices under consideration are anisotropic and highly dispersive. It has been shown in \cite{Belov} that spatial resolutions of $\Delta x<\lambda/80$ are necessary for the FDTD modeling of dispersive left-handed media, in order to avoid spurious resonances which are caused by numerical errors. Moreover, there are discrepancies between the analytical and the numerical values of the material parameters due to the finite spatial grid size used in FDTD simulations. In earlier work on the computational aspects of cloaking devices \cite{Yan}, we have demonstrated that spatial resolution smaller than $\lambda/80$ is also necessary to accurately describe the material parameters in such schemes. Hence, in the work presented in this paper, a very fine spatial resolution of $\Delta x=\lambda/150$ is safely chosen in order to avoid such issues.

Another major challenge in FDTD simulations of dispersive devices based on transformation electromagnetics is late-time numerical instabilities \cite{Argyropoulos}. These instabilities are mainly caused from the strong variations of the spatially varying material parameters. The extreme parameter values, combined with the staircase approximation of the device's cylindrical structure, can lead to spurious cavity resonances, which are visualized as accumulated charges at the interfaces between the boundaries of the device and the surrounding space. In order to mediate these instabilities, we have applied a locally spatial averaging technique \cite{Lee} for the anisotropic field components of the constitutive equations. Finally, corrected values of the plasma and collision frequencies have been used throughout the dispersive simulations, which are taking into account the finite time step of the FDTD technique.

\section{Bandwidth of dispersive cylindrical cloaks}
\label{Bandwidth and Loss Limitation of cylindrical cloaks}

In this section we use the time-domain capabilities of the FDTD technique in order to evaluate the bandwidth of certain cloaking devices. Since dispersive devices need to be constructed from resonant metamaterial elements in order to achieve the designed non-conventional material parameters, the cloaks theoretically operate correctly only for the single frequency for which the parameters of Eqs. (\ref{parametersofcloaking}) or (\ref{parametersofreducedcloaking}) are satisfied. Here we define the cloak's bandwidth through the frequency range over which the transmitted field components are enhanced compared to the corresponding transmitted field components observed when the same excitation input pulse is impinging on a bare non-cloaked PEC cylinder.

The domain of the FDTD simulations can be seen in Fig. \ref{domainandblueshift}(a) and the dimensions of the devices are the same as mentioned in section \ref{Radially-Dependent Dispersive FDTD}. The devices are excited with a plane wave pulse centered at $2$ GHz confined in a broadband Gaussian envelope with a FWHM bandwidth of $1$ GHz. The magnetic field values $H_{z}$ are spatially averaged along the parallel to the x-axis line segment $L_{2}$, approximately $2.8 \lambda$ away from the device's core. The transmitted spectrum is then retrieved from the time-dependence of the averaged field signals, which is next divided by the spectrum of the input pulse, yielding the transmission amplitude as a function of the frequency.

\begin{figure}[t]
\centering
\includegraphics[width=8.0cm]{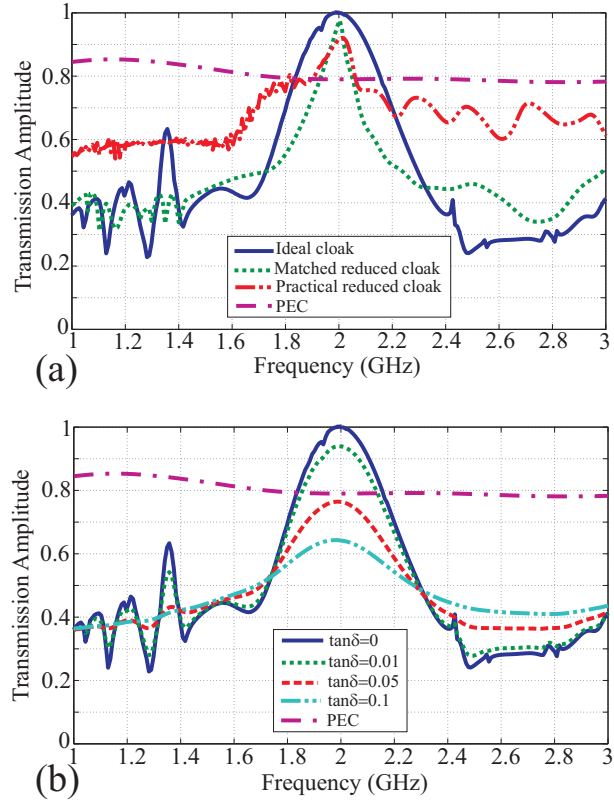}
\caption{(a) The transmission amplitude of the ideal, matched reduced and practical reduced cylindrical cloak. The transmission of a bare PEC cylinder is also plotted to calculate the cloak's effective bandwidth. (b) The transmission amplitude of the ideal cylindrical cloak when different losses are introduced.}
\label{s21parameters}
\end{figure}

The bandwidth performance of the ideal cylindrical cloak \cite{Cummer}, the matched reduced cylindrical cloak \cite{Qiu} and the practical reduced cylindrical cloak \cite{Cai} are compared to the transmission amplitude when a bare PEC cylinder is illuminated. The latter provides a reference point as we assume that a cloak operates only when it enhances the transmission amplitude compared to the bare cylinder case. The comparison between the transmission amplitudes of the different cloaking devices is shown in Fig. \ref{s21parameters}(a).

We observe that the ideal cloak exhibits a transmission amplitude of $1$ at the operating frequency of $2$ GHz, when the corresponding value in the case of the bare cylinder is approximately $0.8$. The calculated effective bandwidth is $11.5\%$. Among the proposed cloaks, this is the most broadband device as the bandwidths of the practical reduced cloak and the matched reduced cloak are only $4.6\%$ and $2.5\%$, respectively. Thus, the approximate nature of the latter two devices significantly impacts its bandwidth performance. As expected, the transmission amplitude of both approximate devices is lower compared to the amplitude of the ideal cloak. However, the scattering performance of the matched reduced cloak is better compared to the practical reduced cloak, because its design allows it to be matched to the surrounding free space. Thus, there is a trade-off between the bandwidth and the scattering performance for the two approximate cloaking designs.

As a final note, the cloaking bandwidth depends on the exact design of the cloak and its corresponding material parameters. For example, the PEC cylinder in these scenarios can be covered with thinner or thicker cloaks, other than the choices we have made here. Thinner cloaks require more extreme material parameters, as indicated by Eqs. (\ref{parametersofcloaking}) and (\ref{parametersofreducedcloaking}), and are thus expected to be even more narrowband. On the other hand, by increasing the cloak's dimensions, its effective bandwidth is also expected to increase.

One of the major practical issues with metamaterial-based devices is the effect of the losses that are inherent in the dispersive resonating elements required to obtain the design material parameters \cite{podolskiy2005nss}, which may inhibit their proper operation. Here we test the effect of losses in the cloaking bandwidth of the ideal cylindrical cloak. Similar as earlier in this section, FDTD simulations are performed with identical simulation domain and dimensions of the structure. Losses are introduced in the FDTD code in the same way as in our previous works \cite{Argyropoulos,argyropoulos2008mle}, i.e. by including a complex component $-j\tan\delta$ into the material parameters, which is increased up to $0.1$.

The transmission amplitude of the ideal cylindrical cloak as a function of frequency for various loss tangents can be seen in Fig. \ref{s21parameters}(b). We observe that the performance of the device is significantly affected with the increase of the losses, as higher losses decrease the transmission amplitude as well as the bandwidth of the cloak. It is interesting that if $\tan\delta\geq0.05$, which is a typical loss value for metamaterial structures close to resonance, then the performance of the ideal cloak is worst than a bare PEC cylinder and the cloaking phenomenon effectively ceases to exist. Similar results are observed when losses are introduced into the approximate cloaking structures (not shown), which implies that a practical, real world cloak will eventually require very low-loss metamaterial elements in order to operate properly.

\section{Spectral Response of the Ideal Cylindrical Cloak}
\label{Spectral Response of the Ideal Cylindrical Cloak}

\begin{figure}[h]
\centering
\includegraphics[width=8.0cm]{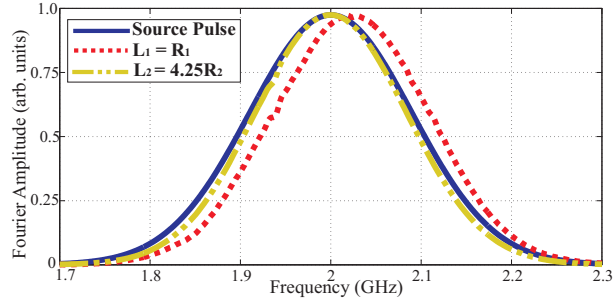}
\caption{Blueshift effect observed in the normalized frequency spectra of transmitted Gaussian narrowband pulses through the ideal cylindrical cloak for two averaging line segments $L_{1}$ and $L_{2}$ of Fig. \ref{domainandblueshift}(a). The line segments are on the same position on the y-axis, $1.5 \lambda$ away from the cloak's outer shell, but have different lengths along the x-axis ($L_{1}=R_{1}$, $L_{2}=4.25R_{2}$). The effect is stronger for the line segment $L_{1}$, near the center of the cloaking structure.}
\label{blueshiftalone}
\end{figure}

In this section, the spectral response of the lossless ideal cylindrical cloak is investigated. The ideal cloak is excited with a narrowband Gaussian pulse with a bandwidth of $200$ MHz (FWHM), centered at a frequency of $f_{0}=2.0$ GHz. The pulse is chosen to be narrowband in order to pass largely unaffected through the ideal cloak's allowed bandwidth (see Fig. \ref{s21parameters}(a)). The results of the spectral content of the transmitted pulses recorded on each line segment of Fig. \ref{domainandblueshift}(a) are seen in Fig. \ref{blueshiftalone}, which are compared to the spectrum of the incident pulse.

We observe that the central frequency of the transmitted pulse is shifted significantly only for the averaged field values corresponding to the shorter line segment $L_{1}$, where the field distribution is strongly affected from the presence of the cylindrical cloak. When averaging over the longer segment $L_{2}$ most of the pulse is transmitted unaffected and, as a result, the frequency shift effect is not observed. Only a small narrowing of the pulse is detected when averaging over $L_{2}$, which is a result of the finite bandwidth of the device (Fig. \ref{s21parameters}(a)). For this simulation scenario, the blueshift is calculated to be $\Delta f=22.5$ MHz, which is $1.1\%$ deviation from the central frequency of $2$ GHz. Hence, an instrument which is capable of resolving spectral deviations smaller than $1.1\%$, will detect the presence of the cloak.
\begin{figure}[b]
\centering
\includegraphics[width=9.0cm]{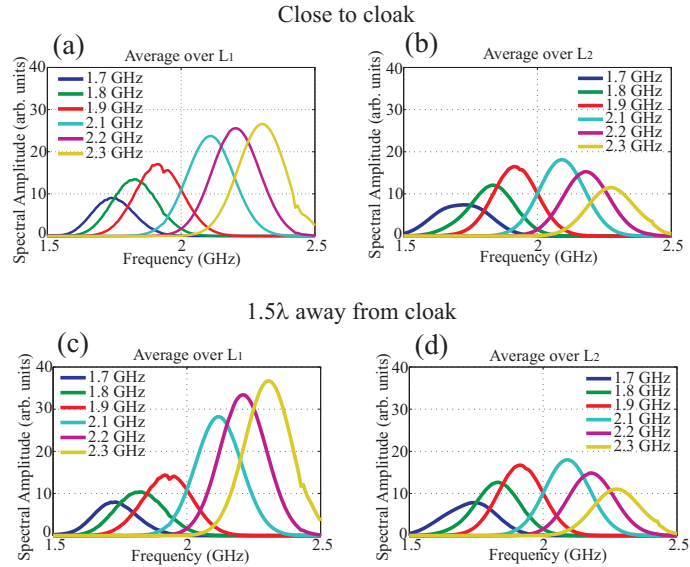}
\caption{Frequency spectra of six transmitted narrowband pulses, centered at different frequencies, after impinging on the ideal cloak. The cloak's design frequency is $2$ GHz. The fields are averaged over two different line segments $L_{1}$ ((a), (c)) and $L_{2}$ ((b), (d)) which are perpendicular to the direction of propagation as shown in Fig. \ref{domainandblueshift}(a). The line segments are positioned either right after the cloak's outer boundary ((a), (b)) or at a distance $1.5\lambda$ away ((c), (d)).}
\label{transmittedpulses}
\end{figure}
\begin{figure*}
\centering
\includegraphics[width=15.0cm]{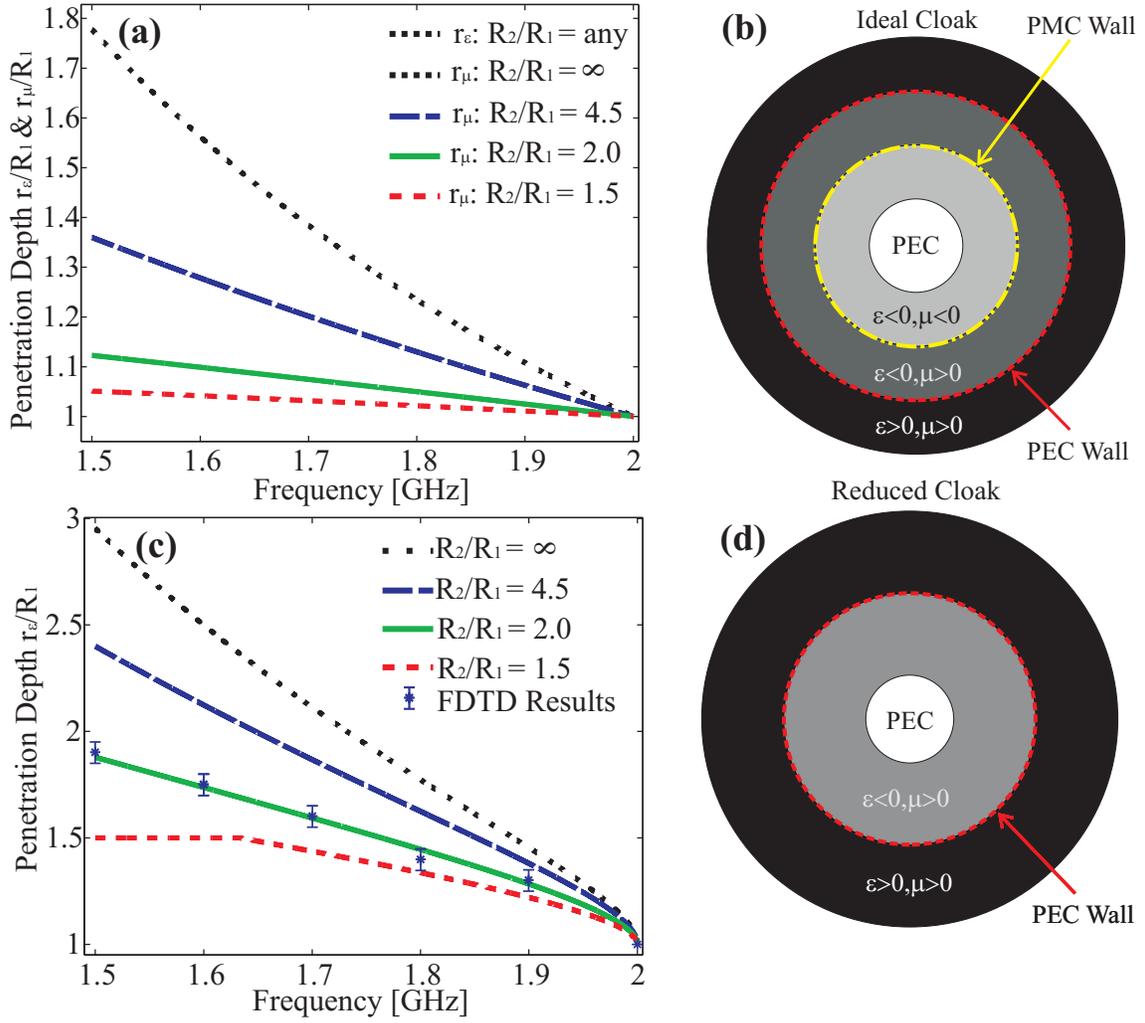}
\caption{(a) Analytical normalized penetration depths for ideal cylindrical cloaks with different dimensions. (b) Regions formed inside the ideal cylindrical cloak, when it operates at $f<f_{0}$ frequencies. (c) Analytical normalized penetration depths for matched reduced cylindrical cloaks with different dimensions. The numerically estimated penetration depths are also shown for a cloak with thickness $2\lambda/3$. The error bars indicate numerical uncertainty on the field cutoff radius. (d) Regions formed inside the matched reduced cylindrical cloak, when it operates at $f<f_{0}$ frequencies.} \label{penetrandregionsbothcloaks}
\end{figure*}

To further investigate the blueshift effect, six narrowband Gaussian pulses are independently launched towards the ideal $2$ GHz cloak, at central frequencies of $1.7$, $1.8$, $1.9$, $2.1$, $2.2$, $2.3$ GHz respectively. The FWHM bandwidth of each pulse is $200$ MHz. The spectral distribution of the averaged field values of the transmitted pulses along the line segments $L_{1}$ and $L_{2}$ depicted in Fig. \ref{domainandblueshift}(a) $1.5\lambda$ away of the cloak is once again evaluated, and is shown in Figs. \ref{transmittedpulses}(c), (d). In addition, we also record the transmitted spectral distribution along two new line segments of the same length $L_{1}$ and $L_{2}$ using the same excitation signals, but this time the segments are positioned immediately after the cloaking structure. These results are shown in Figs. \ref{transmittedpulses}(a), (b).

Firstly, it is observed that when the fields are averaged over the $L_{2}$ line (Figs. \ref{transmittedpulses}(b), (d)) the differences between the pulses above and below the cloak's central frequency are small. The spectrum is affected symmetrically around the central frequency because of contributions from field energy away from the device that does not interact with the cloak. The lower amplitudes recorded away from the central frequency region are caused by the overall finite bandwidth of the cloak, as shown in Fig. \ref{s21parameters}(a). Secondly, when recording over the short line segment $L_{1}$ (Figs. \ref{transmittedpulses}(a), (c)), it is observed that the pulses with central frequencies greater than the cloak's operating frequency ($f>f_{0}$) maintain their overall shape but are reinforced in amplitude. On the contrary, the pulses with central frequencies less than the cloak's operating frequency ($f<f_{0}$) are dissipated, when they are passing through the structure. This behavior gives rise to the frequency shift effect observed in Fig. \ref{blueshiftalone}. The effect is stronger away from the cloak (Fig. \ref{transmittedpulses}(c)), as opposed to near the cloak (Fig. \ref{transmittedpulses}(a)), since the field has recombined behind the device in the former case. At an even greater distance away from the device, the amplitudes of the transmitted pulses do not evolve significantly.

We now investigate the effective structure that an off-frequency incident wave perceives inside the dispersive cloaking shell. Depending on the frequency of the incident wave, the material parameters of the ideal cloak (Eq. (\ref{parametersofcloaking})) may become equal to zero somewhere inside the cloaking shell. These locations are then effectively acting as a PEC wall (when $\varepsilon_{r}=0$) and perfect magnetic conductor (PMC) wall (when $\mu_{z}=0$) for that frequency, beyond which the fields theoretically do not penetrate \cite{zhang2008rab}. The radial locations $r_{\varepsilon}$, $r_{\mu}$, where the permittivity and permeability respectively vanish for the ideal cloak, are given as a function of frequency by (using Eqs. (\ref{radandfreqdependparam_e}), (\ref{radandfreqdependparam_m}), (\ref{parametersofcloaking})):
\begin{eqnarray}
r_{\varepsilon}(\omega<\omega_{0})&=&R_{1}\frac{\omega_{0}^2}{\omega^2}\\
\label{electricpendepth}
r_{\mu}(\omega<\omega_{0})&=&\frac{R_{1}}{1-\left(1-\frac{\omega^2}{\omega_{0}^2}\right)\left(1-\frac{R_{1}}{R_{2}}\right)^2}
\label{magneticpendepth}
\end{eqnarray}

The theoretical field penetration depths normalized to the object's radius $R_{1}$ are plotted in Fig. \ref{penetrandregionsbothcloaks}(a) for different dimensions of the cloak as a function of frequency, for $f<f_{0}$ (for $f>f_{0}$ no such walls are formed). It is observed that always $r_{\varepsilon}>r_{\mu}$, with $r_{\mu}\rightarrow r_{\varepsilon}$ for any $\omega$ as $R_{2}\gg R_{1}$. This implies that a PEC wall will form first as the wave approaches the core of the device, beyond which $\varepsilon_{r}<0$. If a material with positive index of refraction was formed between that wall and the device's PEC core, no fields would penetrate further inside. However, further inside the cloak a PMC wall will be then formed, beyond which also $\mu_{z}<0$. Thus, a negative index shell will be formed for some frequencies near the core of the device, which can trigger phenomena such as superscattering \cite{yang2008ses}. This three-layer effect is unique to the ideal cylindrical cloak (for TE or TM polarizations) because two material parameters can become simultaneously negative due to dispersion, something that is not occurring in either the ideal spherical cloak or the matched reduced cylindrical cloak. These layers are depicted graphically in Fig. \ref{penetrandregionsbothcloaks}(b) for $f<f_{0}$ frequencies.

In addition, since for $f=f_{0}$ the material parameters become zero at the inner core $r=R_{1}$, waves with frequencies $f<f_{0}$ will sample negative values of the material parameters near the core, while waves with frequencies $f>f_{0}$ will sample strictly positive material parameters values everywhere in the cloaking shell. Thus, the former waves ($f<f_{0}$) will perceive the device as conducting scatterer, while the latter ($f>f_{0}$) will perceive it as a dielectric and magnetic material, with positive permittivity and permeability.

This latter effect, that waves at different frequencies perceive electromagnetically different devices, is demonstrated by launching three monochromatic plane waves at frequencies $1.7$ GHz, $2.0$ GHz and $2.3$ GHz against the ideal cloak. The real part of the converged magnetic field values are shown in Fig. \ref{reducedcloakfields}(a)-(c). We observe in Fig. \ref{reducedcloakfields}(a) that for the $1.7$ GHz wave, little field reaches the core of the device due to the aforementioned walls that form, and thus a strong shadow is observed behind the device. On the contrary, in Fig. \ref{reducedcloakfields}(c), the $2.3$ GHz wave perceives the cloaking shell as a positive index material and field is distributed even inside the device. As a reference, the field distribution on the nominal frequency of $2.0$ GHz is also shown in Fig. \ref{reducedcloakfields}(b). The positive-to-negative material parameters transitions that occur for $f<f_{0}$ but not for $f>f_{0}$ give rise to the frequency modulation that occurs inside the cloaking shell (shown in Fig. \ref{transmittedpulses}(a), (c)), as waves at $f<f_{0}$ are dissipated due to the scattering from a conductor, while others at $f>f_{0}$ are scattered from a positive-index material. This ultimately produces the frequency shift observed in Fig. \ref{blueshiftalone}.

\section{Spectral Response of the Reduced Cylindrical Cloak}
\label{Spectral Response of the Reduced Cylindrical Cloak}

\begin{figure}[t]
\centering
\includegraphics[width=9.0cm]{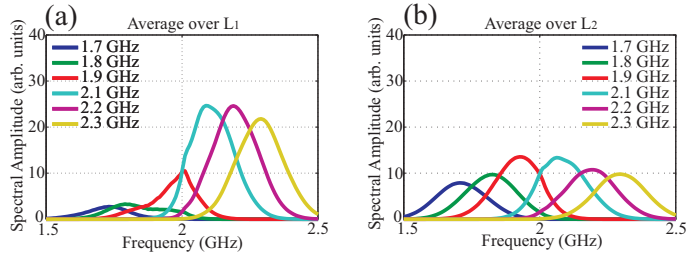}
\caption{Frequency spectra of six transmitted narrowband pulses, centered at different frequencies, after impinging on the matched reduced cloak. The cloak's design frequency is $2$ GHz. The fields are averaged over two different line segments $L_{1}$ (a) and $L_{2}$ (b), as shown in Fig. \ref{domainandblueshift}(a), which are perpendicular to the direction of propagation. The line segments are positioned at a distance $1.5\lambda$ away from the cloak's outer shell.}
\label{transmittedpulsesreduced}
\end{figure}

We now focus on the matched reduced cylindrical cloak \cite{Qiu}, which is easier to be practically implemented compared to the ideal cloak. This material parameter set is more similar to the ideal spherical cloak \cite{Pendry}, in the sense that, for a given polarization, only one parameter is radially-dependent and dispersive ($\varepsilon_{r}$ for TE waves), whereas the others are constant and conventional. FDTD simulations are performed over the same domain as before (Fig. \ref{domainandblueshift}(a)) and the cloak has the same dimensions. Again, narrowband Gaussian pulses at various central frequencies are launched towards the cloak. The transmitted pulses in the frequency domain, after impinging on the cloak, recorded at the two line segments $L_{1}$ and $L_{2}$ of Fig. \ref{domainandblueshift}(a), at a distance $1.5\lambda$ away from the device, are seen in Figs. \ref{transmittedpulsesreduced}(a) and (b), respectively.

The transmitted pulses appear slightly distorted compared to the ideal cloak (Fig. \ref{transmittedpulses}) due to the fact that the reduced cloak examined here is designed to operate approximately. However, the features of the transmitted pulses are similar to the transmission observed through the ideal cloak, where frequencies $f<f_{0}$ are dissipated while frequencies $f>f_{0}$ are enhanced. The effect is once again much stronger when averaging the fields near the cloak's core, and gives rise to a similar frequency blueshift effect.
\begin{figure*}
\centering
\includegraphics[width=15.0cm]{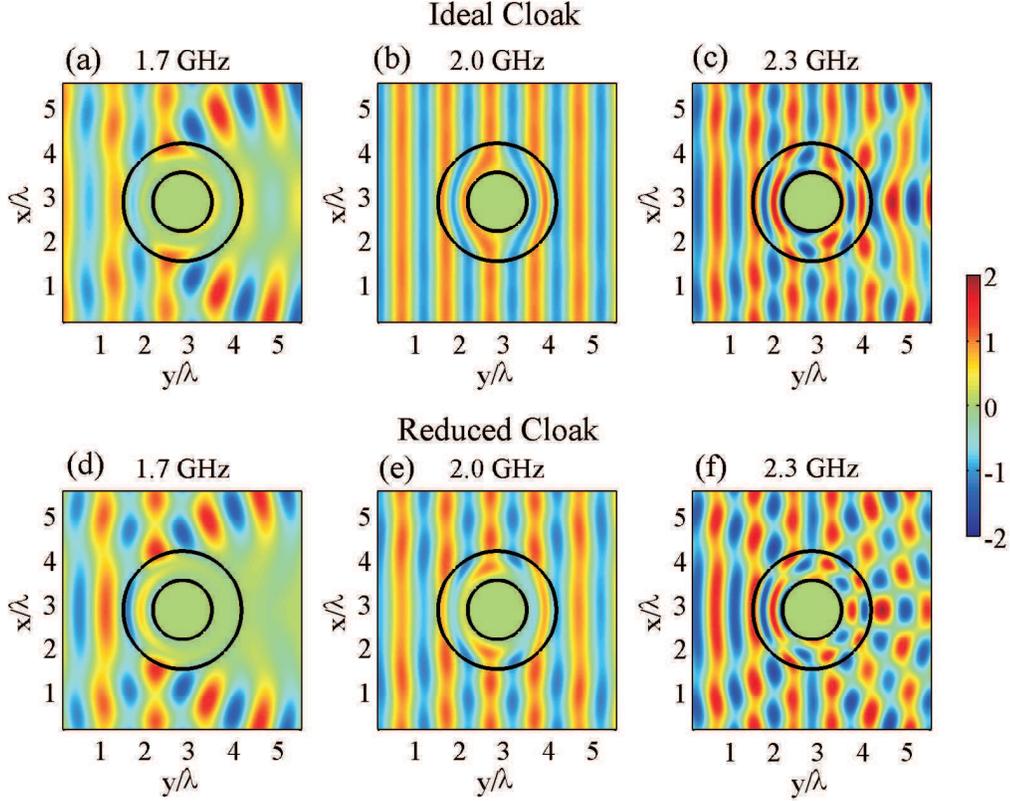}
\caption{Real part of the magnetic field amplitude distributions ($H_{z}$) when plane waves of different frequencies are impinging on the ideal ((a)-(c)) and matched reduced ((d)-(f)) cloak, after steady state is reached. The amplitude scale is normalized such that $1$ is the maximum plane wave field amplitude without any device present. (a),(d): At $1.7$ GHz a PEC wall is formed inside the cloaking shell where fields cannot penetrate. The devices behave as a conductive scatterer and a large shadow is observed behind them. (b),(e): At $2.0$ GHz the devices operate at their nominal frequency. For (e) only, imperfections in the field distributions are inherent in the limitations of the cloak's approximate design. (c),(f): At $2.3$ GHz the cloaking material is perceived as a dielectric scatterer from the incident wave. Thus no significant shadow is formed behind the devices.}
\label{reducedcloakfields}
\end{figure*}

For some frequencies $f<f_{0}$, the dispersive parameter $\hat{\varepsilon}_{r}$ will become equal to zero at a specific location inside the cloaking shell, forming a boundary beyond which wave penetration is very weak. The penetration depth is calculated analytically with a similar procedure as before and it is given by the equation:
\begin{eqnarray}
r_{\varepsilon}(\omega<\omega_{0})=\frac{R_{1}}{1-\sqrt{\left(1-\frac{\omega^2}{\omega_{0}^2}\right)\left(1-\frac{R_{1}}{R_{2}}\right)}}
\label{electricpendepthreducedset}
\end{eqnarray}
It is normalized to $R_{1}$ and plotted as a function of the $f\leq f_{0}$ frequencies in Fig. \ref{penetrandregionsbothcloaks}(c) for cloaks with different thicknesses. Moreover, the locations, where the impenetrable wall is formed, are deduced approximately through the FDTD simulation by identifying the points where the magnetic field $H_{z}$ becomes zero. They are also shown in Fig. \ref{penetrandregionsbothcloaks}(c) for a specific cloak design and are found to be in excellent agreement with the analytical predictions of the penetration depth derived from Eq. (\ref{electricpendepthreducedset}). The layers that form inside the reduced cloak are depicted graphically in Fig. \ref{penetrandregionsbothcloaks}(d) for $f<f_{0}$ frequencies.

Similar to the procedure followed in the previous section for the ideal cloak, in order to explain the different behavior of the pulses above and below the central frequency, along with the formation of the PEC wall, three monochromatic plane waves are launched with frequencies $1.7$, $2.0$ and $2.3$ GHz towards the $2$ GHz matched reduced cloak. The real part of the converged field amplitude values are seen in Figs. \ref{reducedcloakfields}(d)-(f). As a reference, Fig. \ref{reducedcloakfields}(e) shows the field distribution at the cloak's nominal frequency, where the wave energy mostly recomposes smoothly behind the device. The imperfections observed in the wavefronts are due to the approximate nature of the reduced cloak's design.

For the plane wave with frequency $1.7$ GHz, according to Eq. (\ref{electricpendepthreducedset}), an impenetrable wall is expected to form approximately $0.36\lambda$ away from the inner core. This is indeed observed in Fig. \ref{reducedcloakfields}(d), where little field reaches the metallic core. As was discussed previously, due to the PEC wall the wave at that frequency should perceive the device as a conductive scatterer. This is confirmed in the simulation by observing the large shadow that appears behind the device, similar to the shadow that typically appears by the scattering of a plane wave off a conductive cylinder. This shadow is the reason for the reduced amplitude recorded for wave pulses at frequencies $f<f_{0}$, as was pointed out in Fig. \ref{transmittedpulsesreduced}(a). The shadow is similar to the one observed in the case of the ideal cloak, as shown in Fig. \ref{reducedcloakfields}(a).

For the plane wave with frequency $2.3$ GHz, no PEC wall is predicted to exist. Instead, the wave perceives the cloaking shell as a dielectric scatterer, since the relative permittivity values are always larger than zero for $f>f_{0}$. Indeed, the field distribution shown in Fig. \ref{reducedcloakfields}(f) indicates that the fields penetrate all the way into the inner core, and that no shadow appears behind the device. On the contrary, the interference effects cause the field amplitude behind the cloak to be locally larger than the amplitude of the incident field (equal to unity in Fig. \ref{reducedcloakfields}), thus enhancing these $f>f_{0}$ frequencies compared to the $f<f_{0}$ frequencies. This explains the modulation pattern that appears in Figs. \ref{transmittedpulsesreduced}(a) which ultimately causes the blueshift effect presented in Fig. \ref{blueshiftalone}.

\section{Temporal and Spatial Responses of the Ideal Cylindrical Cloak and Concentrator}
\label{Temporal and Spatial Responses of the Ideal Cylindrical Cloak and Concentrator}

\begin{figure*}
\centering
\includegraphics[width=15.0cm]{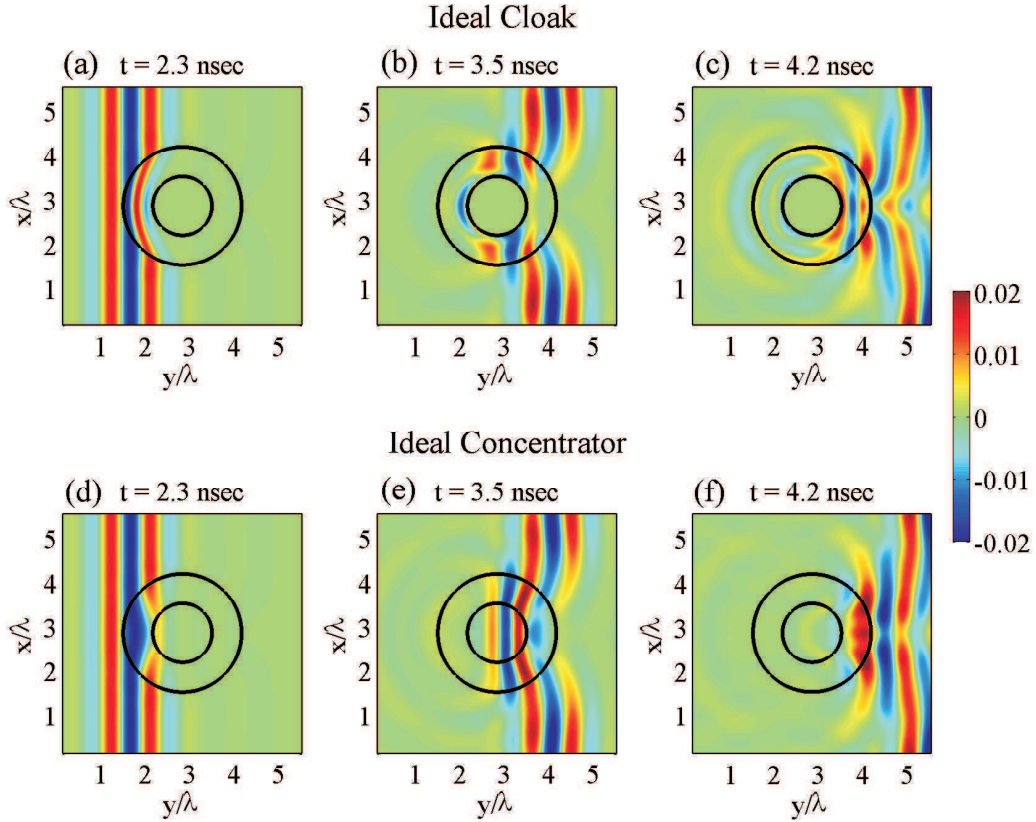}
\caption{Time-dependent snapshots of the real part of the magnetic field amplitude distribution ($H_{z}$) when a $1$ GHz (FWHM) Gaussian pulse is impinging on dispersive devices. (a)-(c): Ideal cloak. (d)-(f): Ideal concentrator. In (a),(d) the pulse has reached the first half of the devices. In (b),(e) the pulse leaves the dispersive regions. The wavefronts near the center of the devices are delayed compared to the wavefronts away from the central regions. In (c),(f) a delay appears near the center of the devices, even though the waves have recomposed. Reflections are observed due to the broad bandwidth of the incident pulse.}
\label{transientresponsescloakconc}
\end{figure*}
Another interesting effect takes place in the ideal cloaks for non-monochromatic waves. While light rays that travel away from the core of the cloak propagate at the speed of light in free space, rays that traverse the core region experience time delays \cite{chen2008tda}, due to the increased values of the permittivities and permeabilities introduced by the anisotropic cloaking materials. This effect is stronger near the center of the cloak, where in theory some material parameters reach infinite (or realistically very large) values. As we shall show, the phenomenon is visible when a temporally finite pulse is incident towards the cloak: wavefronts propagating away from the core reach the other side of the domain (behind the cloak) sooner than wavefronts propagating near the center.

The time-delay effect has been predicted theoretically for a 3D ideal spherical cloak \cite{chen2008tda,zhangoptexpr2009}. In this paper, this dynamic effect is demonstrated and exploited with the FDTD method for the ideal cylindrical cloak. The same dimensions, as before, are chosen for the tested cloak. A broadband Gaussian pulse is launched and is passing through the 2D structure. It has a bandwidth of $1$ GHz, centered at the cloak's nominal frequency of $2$ GHz. The real part of the magnetic field amplitude distribution is calculated as a function of time through the FDTD simulation. Three time snapshots of the pulse are depicted in Figs. \ref{transientresponsescloakconc}(a)-(c), as the pulse is propagating through the device. When the pulse is recomposed after the cloak (Fig. \ref{transientresponsescloakconc}(c)), the wavefronts are not flat anymore due to the experienced time delay, which is more intense closer to the cloak's inner boundary. The group velocity is reduced close to the inner boundary of the structure, whereas the phase velocity is approaching large values.

In this example, the wavefronts near the center of the cloak are delayed by approximately $0.27$ ns compared to wavefronts that propagate mostly undisturbed away from the cloak. The bending of the waves is caused by the extremely high values of the radially-dependent cloaking parameters (Eq. (\ref{parametersofcloaking})) due to the dilation of the domain from a point to a circle ring, the device's inner core. Hence, the time delay of the pulse is inherent to the design of the cloak and, as a result, unavoidable. Thus, one could take advantage of this effect in order to detect the presence of the cloaking device with an appropriate instrument. It is present in both the ideal and reduced (not shown here) 2D cylindrical cloaks. It is also the main reason for the long time required for the FDTD simulations to reach steady-state results \cite{Liang} in such simulations.

The time-delay effect is not limited to cloaks, but generally appears whenever large variations of the values of the material parameters are required for a given device. To illustrate a different scenario, Figs. \ref{transientresponsescloakconc}(d)-(f) show time snapshots as the same broadband Gaussian pulse is impinging on the ideal concentrator \cite{Rahm}, a device designed to focus electromagnetic fields inside a small region of space ($r<R_{1}$ in this case). The dimensions of the device are the same as the cloak's discussed earlier. The time-delay effect is observed after the wave recomposes behind the device, especially near the core, where the wavefronts are delayed by roughly $0.18$ ns compared to the unperturbed waves. The effect is weaker compared to the cloak, since none of the material parameters reach extreme values (maximum value of permittivities and permeability is $3$) \cite{Rahm}. Meanwhile, the phase and group velocities have comparable values. It should also be noted that reflected waves emerge as the pulses impinge on either device (Figs. \ref{transientresponsescloakconc}(c) and (f)). This is a result of the large number of frequency components contained in the Gaussian pulses, since the devices can only operate properly at their nominal frequency of $2$ GHz.
\begin{figure}[t]
\centering
\includegraphics[width=7.0cm]{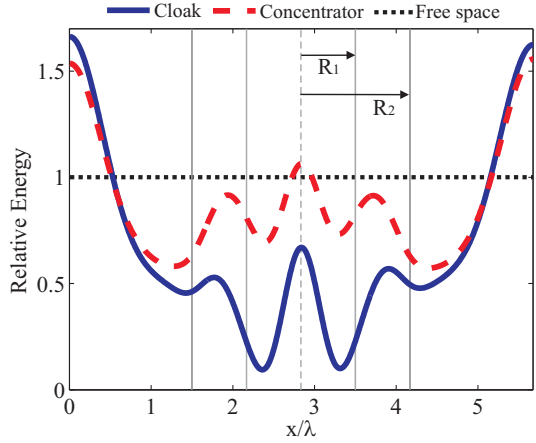}
\caption{Relative accumulated energy distribution for ideal cloak and concentrator. In the amplitude scale shown, $1$ is the accumulated energy distribution of free space propagation. The lines $R_{1}$ and $R_{2}$ indicate the inner and outer radii of the devices. The distributions are not perfectly symmetric because of the finite discretization used in the FDTD simulations.}
\label{energyoutofcloakconc}
\end{figure}

In addition to the temporal disturbances induced on the Gaussian pulses by the dispersive devices, their spatial distribution is affected as well \cite{zhangoptexpr2009}. To illustrate this effect, the time-integrated energy distribution crossing the line segment $L_{2}$ (Fig. \ref{domainandblueshift}(a)) for both ideal cloak and concentrator is recorded in the setup of Fig. \ref{transientresponsescloakconc}, and it is shown in Fig. \ref{energyoutofcloakconc}. We observe that the energy of the non-monochromatic source is not distributed uniformly outside of the devices due to scattering of the frequency components away from $f_{0}$. The cloak has more extreme parameters than the concentrator, which leads to less accumulated energy inside the device and more scattered energy outside. It is interesting that the energy reaching the center of both structures is lower than the scattered energy near the edges of the simulation domain. Hence, the energy is not properly spatially recombined behind either device.

\section{Conclusions}
\label{Conclusions}
Concluding, the properties of the metamaterial dispersive devices under non-monochromatic illumination are studied. The inherent dispersive nature of the devices is the main reason behind the transient effects observed, which is affecting their broadband performance. We investigate the origin of the frequency shifts that appear in both the ideal and matched reduced cylindrical cloaks through dispersive FDTD simulations, using narrowband Gaussian pulses as illumination. In addition, the time-delay and spatial non-uniformity effects that occur near the center of the dispersive regions of the ideal cloak as well as the ideal concentrator are demonstrated. These are important effects that should be taken into account in future applications of dispersive metamaterial devices.

The ideal cylindrical cloak and concentrator, work properly only for monochromatic wave incidence. However, they become non practical when they are excited with non-monochromatic radiation, which is the type of electromagnetic radiation found in nature. The effect of losses is also significant for such dispersive devices. In general, the transmitted signal, passing through devices derived from coordinate transformations, will have its spectral, temporal and spatial profiles altered. These restraints could be avoided if the cloak is constructed from broadband active metamaterials \cite{tretyakov2001mmw,chen2006atm}. Alternatively, novel non-dispersive devices based on conventional materials \cite{Kallos} need to be implemented.

\bibliography{referencesblueshift}

\end{document}